\documentclass[sn-nature]{sn-jnl}

\usepackage{graphicx}%
\usepackage{multirow}%
\usepackage{amsmath,amssymb,amsfonts}%
\usepackage{mathrsfs}%
\usepackage[title]{appendix}%
\usepackage{xcolor}%
\usepackage{textcomp}%
\usepackage{manyfoot}%
\usepackage{booktabs}%
\usepackage{algorithm}%
\usepackage{algorithmicx}%
\usepackage{algpseudocode}%
\usepackage{listings}%

\newcommand{\uvec}[1]{\boldsymbol{\hat{\textbf{#1}}}}
\begin{document}

\title[Article Title]{Off-Resonant Detection of Domain Wall Oscillations Using Deterministically Placed Nanodiamonds}


\author[1]{\fnm{Jeffrey} \sur{Rable}}

\author[1]{\fnm{Jyotirmay} \sur{Dwivedi}}

\author*[1,2]{\fnm{Nitin} \sur{Samarth}}\email{nsamarth@psu.edu}

\affil*[1]{\orgdiv{Department of Physics}, \orgname{The Pennsylvania State University}, \city{University Park}, \postcode{16802}, \state{PA}, \country{USA}}

\affil[2]{\orgdiv{Department of Materials Science \& Engineering}, \orgname{The Pennsylvania State University}, \city{University Park}, \postcode{16802}, \state{PA}, \country{USA}}



\abstract{Nitrogen-vacancy (NV) centers in diamond offer a sensitive method of measuring the spatially localized dynamics of magnetization and associated spin textures in ferromagnetic materials. We use NV centers in a deterministically positioned nanodiamond to demonstrate off-resonant detection of GHz-scale microwave field driven oscillations of a single domain wall (DW).  The technique exploits the enhanced relaxation of NV center spins due to the broadband stray fields generated by an oscillating DW pinned at an engineered defect in a lithographically patterned ferromagnetic nanowire. Discrepancies between the observed DW oscillation frequency and predictions from micromagnetic simulations suggest extreme sensitivity of DW dynamics to patterning imperfections such as edge roughness. These experiments and simulations identify potential pathways toward quantum spintronic devices that exploit current driven DWs as nanoscale microwave generators for qubit control, greatly increasing the driving field at an NV center and thus drastically reducing the $\pi$ pulse time.}

\keywords{ferromagnetic domain wall, NV center}



\maketitle

\section{Introduction}\label{sec1}

GHz-scale ferromagnetic dynamics, such as spin waves, vortex gyrations, and domain wall oscillations, offer both a fascinating target for quantum sensing and a potential qubit driving source in quantum computing applications. One of these qubits, the nitrogen-vacancy (NV) center in diamond, has been previously used to detect ferromagnetic resonance\cite{Wolfe2014,Page2019,Du2017,Rable2022}, spin wave scattering\cite{Zhou2021}, spin-wave frequency multiplication \cite{Koerner2022}, and vortex gyrations\cite{Trimble2021}. Ferromagnetic dynamics have also been explored as potential sources of microwave driving for NV centers: NV-magnon coupling was first reported in 2017\cite{Andrich2017}, then electrically tuned in subsequent works \cite{Wang2020,Yan2022}, and vortices have been shown to enhance NV spin rotation during qubit addressibility experiments\cite{Wolf2016}. 

Domain walls (DWs) should offer similar advantages to vortices for addressability since they produce large stray fields and they can be moved electrically. Pinned DWs act like a mass in a potential well, which is determined by the ferromagnet's properties; they can oscillate in the GHz range, making them useful for direct NV driving. They can be pinned to specific sites via geometric patterning of the magnetic feature, and can be driven via external microwave fields\cite{Galkiewicz2014}, DC current\cite{Demiray2015,Sbiaa2018}, AC current\cite{Saitoh2004}, and spin transfer torques\cite{Kumar2020}. However, the literature on DW sensing using NVs has been largely focused on characterizing their static structures.\cite{Tetienne2015,Finco2021,Jenkins2019}

\begin{figure}[b]
\includegraphics [width=0.8\textwidth] {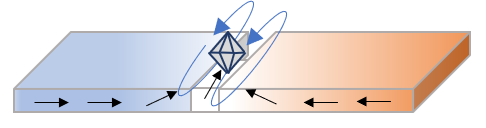}
\caption{\label{fig:Cartoon}Schematic of the measurement concept; an NV center containing nanodiamond is placed over a geometrically-determined DW pinning site in a ferromagnetic nanowire.}
\end{figure}

In this paper, we take the first steps towards NV-DW coupling by off-resonantly detecting GHz-scale transverse domain wall dynamics in NiFe (Permalloy, Py) nanowires using deterministically placed nanodiamonds at a DW pinning site (Fig.~\ref{fig:Cartoon}). We demonstrate that, similar to previous ferromagnetic resonance and vortex gyration measurements \cite{Trimble2021}, driving of domain wall oscillations generates magnetic fields that overlap the NV ground state transition frequency, reducing NV polarized state lifetimes and producing an optical contrast.

\subsection{\label{sec:level2}Domain Wall Oscillation Measurement}


\section{\label{sec:level3}Results}
\begin{figure}
\includegraphics [width=0.8\textwidth] {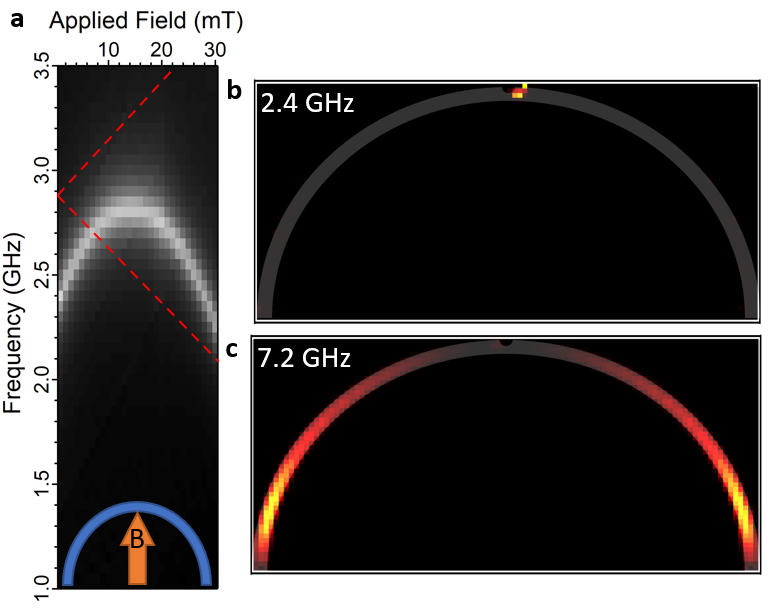}
\caption{\label{fig:Simulations}(a) Simulations showing the frequency of DW oscillations over applied fields in a 300 nm wide bowed nanowire with a 120 nm radius half-circle defect. The dotted red lines denote the NV ground state transition for a field perfectly aligned along the NV axis (b) Spatially resolved image of the mode in (a) at 1 mT. (c) Spatially resolved image of a higher frequency mode at 1 mT, showing that more uniform dynamics occur at much higher frequencies than DW dynamics.}
\end{figure}

Initial simulations with external microwave field driving allow us to find an ideal nanowire geometry for DW detection and NV-DW coupling, with the goal of producing oscillations that would be easily detectable and overlap with the NV ground state transition at an applied field below the DW nucleation field. This would allow us to independently detect the DW away from an NV transition, tune the oscillation frequency and NV ground state transitions into resonance for coupling measurements, and perform control measurements for both these experiments by denucleating the DW. These results led to the fabrication and study of nanowires with widths in the 100's of nm range, with pinning sites composed of a half-circle defect with a radius of approximately 40\% the wire width, most of which should have enabled coupling. For example, the simulation results shown in Fig.~\ref{fig:Simulations} suggest that, for a 300 nm wide wire with a 120 nm defect, DW oscillations should be detectable at about 300 MHz below the NV ground state transition in the absence of an applied field, and that the two should come into resonance at a field below 10 mT. Simulations on the wire initialized with no DW show that one will nucleate at 11 mT applied field.  

\begin{figure}
\includegraphics [width=0.8\textwidth]{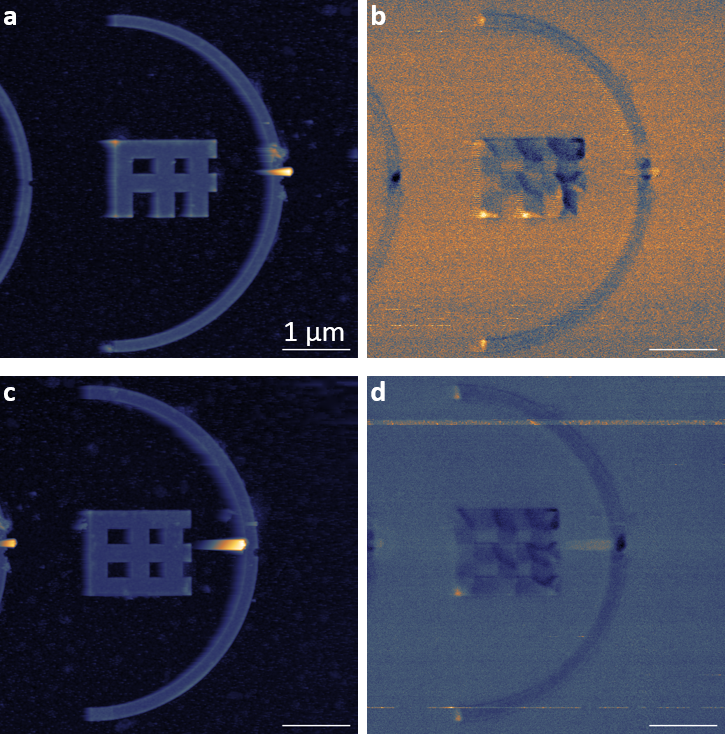}
\caption{\label{fig:NDPlacement}
(a) Atomic force microscopy (AFM) image of nanodiamond placed at the pinning site (bright spot) in a 300 nm wide wire with a 120 nm radius half-circle defect. (b) MFM image of the same site, showing the presence of a DW (dark spot) near the nanodiamond and pinning site. The DW is partly obscured by the nanodiamond, which leaves a small bright streak on the image. (c) AFM image of nanodiamond placed at the pinning site in a 350 nm wide wire with a 140 nm radius half-circle defect. (d) MFM image of the same site, showing the presence of a DW in the wire.}
\end{figure}

\begin{figure}
\includegraphics [width=0.8\textwidth] {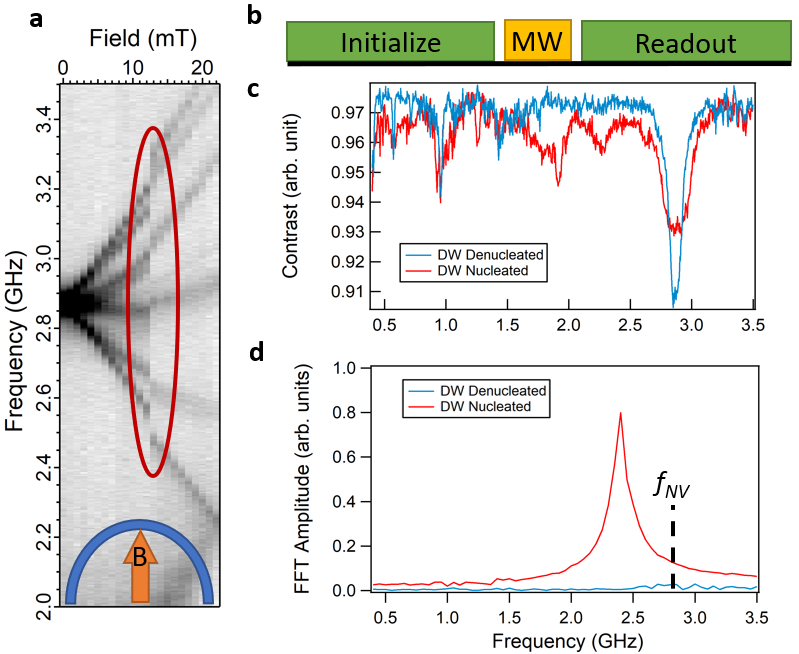}
\caption{\label{fig:Measurements300nm}(a) Continuous-wave electron spin resonance (ESR) measurement performed on a nanodiamond positioned over the pinning site of a 300 nm wide nanowire with a 120 nm defect. The circled discontinuity marks where the DW nucleated. (b) Diagram of the $T_1$ spectroscopy measurement used to detect DW dynamics; green blocks denote the 532 nm laser. (c) $T_1$ spectroscopy measurements of the nanodiamond from (a) performed both with a nucleated domain wall and without in minimal applied field (0.7 mT). (d) Normalized linecut of the simulation results from Fig. \ref{fig:Simulations} (a) at 1 mT, along with the normalized results of a simulation in the same wire without a DW nucleated.}
\end{figure}

We begin our measurements on a 300 nm wide nanowire with a 120 nm wide defect where a nanodiamond is placed over the pinning site (Fig.~\ref{fig:NDPlacement} (a) and (b)) using a pick-and-place protocol reported earlier \cite{Rable2022} and in Methods. Using the nanodiamond located there, we begin measuring the DW nucleation field by monitoring the frequency of the NV ground state transition over increasing applied field; because the Zeeman splitting is approximately 28 GHz/T $\vec{B} \cdot \uvec{n}$, where $\vec{B}$ is the applied field and $\uvec{n}$ is the NV axis unit vector, we can detect the nucleation and the accompanying increase in stray field at the pinning site via the detection of a discontinuity in the ground state transition. This is seen in Fig.~\ref{fig:Measurements300nm} (a) at 13 mT, slightly off from the predicted 11 mT from our simulations. Next, we search for DW oscillations using a $T_1$ spectroscopy measurement (Fig.~\ref{fig:Measurements300nm} (b)). In this measurement, we use the 532 nm laser to polarize the NV into the $m_s = 0$ state, then apply microwaves for 5 \textmu s with the laser off, and turn the laser back on to read out the state during the first 500 ns it is on (signal measurement). A reference measurement is then taken 15 \textmu s later, after the NV spin is re-polarized, with our result being the contrast between the signal and reference measurements. Sweeping across a 400 MHz to 3.5 GHz range both with and without a DW nucleated, this measurement, shown in Fig.~\ref{fig:Measurements300nm} (c), yields a broader range of dynamics than expected in simulations (Fig.~\ref{fig:Measurements300nm} (d)). Rather than a single broad peak centered at 2.4 GHz, we find two broad peaks at approximately 1.8 and 2.3 GHz, with an additional, sharper peak at 1.9 GHz. Earlier studies have suggested that DW dynamics in nanowires are extremely sensitive to edge roughness\cite{Galkiewicz2014}. This leads us to hypothesize that the difference between experimental and simulated results is likely due to the deviation of the shape of the fabricated wires from the ideal design used in our simulations. Furthermore, we find that the signal in the $T_1$ spectroscopy measurement fades rapidly with increasing field, preventing us from realizing resonant coupling (data not shown). At minimal applied field, directly driving the NV ground state transition through Rabi oscillations with and without a DW fails to demonstrate DW-mediated power enhancement at this site; we also do not see any signs of DW-mediated power broadening in our ESR signals (data not shown).

\begin{figure}
\includegraphics [width=0.8\textwidth] {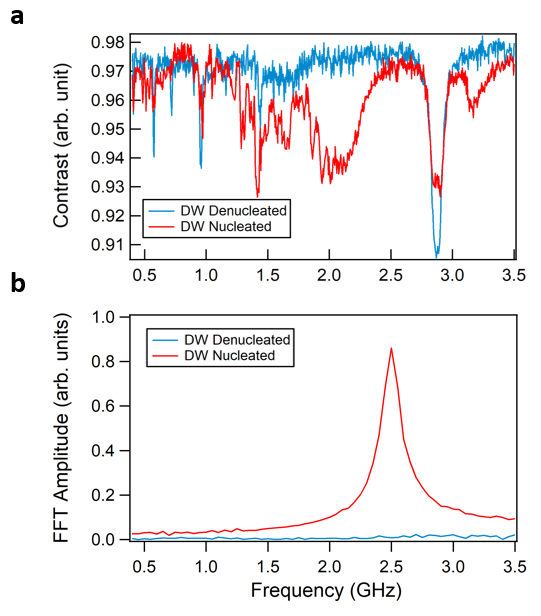} 
\caption{\label{fig:Measurements350nm}(a) $T_1$ spectroscopy measurement on the nanodiamond in Fig. \ref{fig:NDPlacement} (c), located on top of a 350 nm wide wire with a 140 nm radius semicircular defect. (b) Normalized simulations of dynamics in a 350 nm wide bowed wire with a 140 nm radius semicircular defect with and without a DW.}
\end{figure}

\begin{figure}
\includegraphics [width=0.8\textwidth]{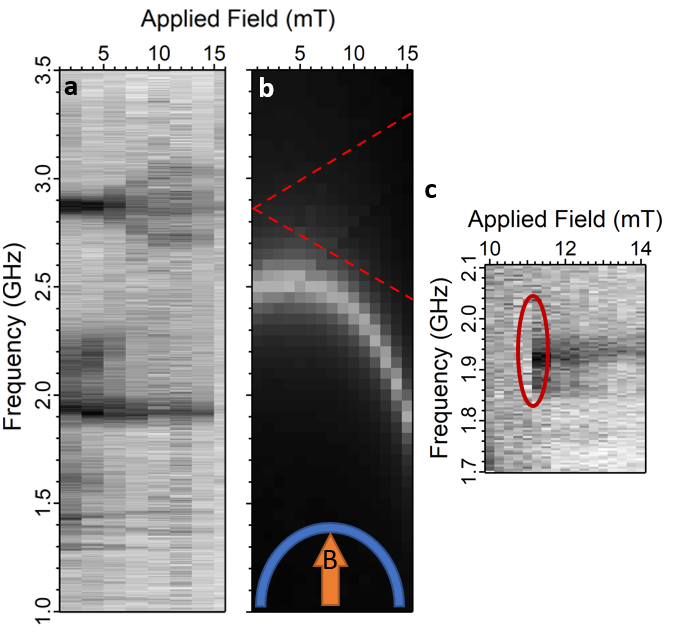}
\caption{\label{fig:Measurements350nmField}(a) $T_1$ spectroscopy measurement performed across different applied fields on the nanodiamond in Fig. \ref{fig:NDPlacement} (c), placed on top of a 350 nm wide wire with a 140 nm radius semicircular defect. (b) Simulations of the DW dynamics of the wire measured in (a); the dashed red line represents the NV ground state transition frequency for an applied field oriented perfectly along the NV axis. (c) Measurement of DW nucleation in the 350 nm wire using the 1.9 GHz peak}
\end{figure}

We then repeat these DW oscillation measurements using $T_1$ spectroscopy on a second, 350 nm wide wire with a 140 nm defect, shown in Figs.~\ref{fig:NDPlacement} (c), with the results shown in Fig.~\ref{fig:Measurements350nm} (a) and Fig.~\ref{fig:Measurements350nmField} (a). This wire does not appear ideal for coupling in simulations (Fig.~\ref{fig:Measurements350nmField} (b)), but is the largest in our series of wires and, as such, should produce the largest stray fields, making it a useful system to confirm our previous results. Similar to the 300 nm wide wire, we see the strongest signal at approximately 2 GHz. However, we also detect some additional peaks near 1.4 GHz and 3.2 GHz; this result is somewhat surprising considering that the 350 nm wire appears to have fewer fabrication errors (Fig. \ref{fig:NDPlacement} (a) and (c)). Fig.~\ref{fig:Measurements350nmField} (a) shows that upon sweeping the applied magnetic field, these additional peaks rapidly fade, and the feature at 2 GHz splits into two distinct features - one that increases in frequency, but fades rapidly, and one that remains constant at 1.9 GHz; neither of these match our simulated domain wall frequency dispersion. This persistent, constant signal at 1.9 GHz, allows us to detect the nucleation field using the domain wall dynamics, rather than the static stray field, demonstrated in \ref{fig:Measurements350nmField} (c); we find that the nucleation field is 11.25 mT, matching our simulations that show nucleation occurs between 11 and 12 mT. Thus, while our dynamic simulations fail to accurately predict DW oscillation frequency, our quasi-static simulations accurately predict the nucleation field.

To elucidate why NV centers can detect DW oscillations off-resonantly, we perform micromagnetic simulations where the DW is driven by a constant sinusoidal driving field and compute the stray field at the NV location. To determine appropriate driving field amplitudes to use in these simulations, we perform Rabi oscillation measurements on our NV at the pinning sites; these results give us a maximum driving amplitude of 1.1 mT at 2.87 GHz, which we take as an approximation of our 1.9 GHz domain wall driving field. For our simulated wire, we measure the generated field at 3.3 GHz to replicate the frequency difference between our measured DW oscillations and NV center zero field splitting, rather than looking directly at the ZFS. We find that the Fourier components at this frequency reach magnitudes of nearly 0.1 mT at the driving field amplitudes we achieved in our experiment, which is sufficient to explain the measured relaxation enhancement.

\begin{figure}[b]
\includegraphics [width=0.8\textwidth] {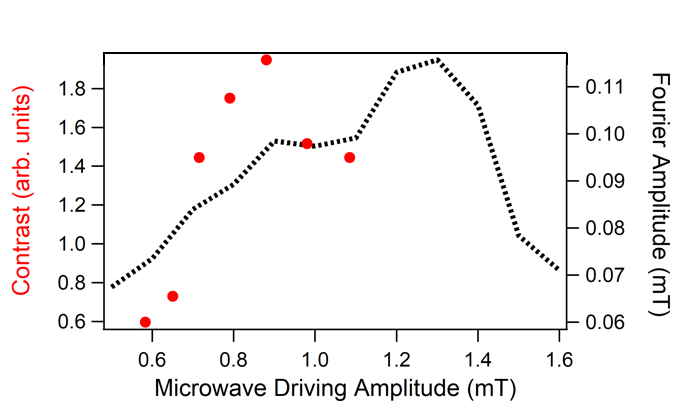}
\caption{\label{fig:Power}NV center DW oscillation peak contrast (red dots) in the 300 nm wide wire and the corresponding simulated 3.3 GHz Fourier component magnitude (dashed line) versus microwave driving field amplitude. The simulated DW stray field used to perform this analysis was measured at 50 nm above the center of the wire, where the pinning site was located.}
\end{figure}

Furthermore, comparing the 3.3 GHz Fourier component to our measured domain wall oscillation peak contrast, a rough proxy for stray field amplitude, in Fig.~\ref{fig:Power}, we can see a similar pattern between the experimental and simulated results, where the stray field amplitude peaks and then drops off substantially. Two possibilities explain the difference in peak locations. First, there are inherent discrepancies between our measured and simulated wires, which we also believe caused the mismatch between observed and predicted DW oscillation frequencies. Second, we measured microwave driving power at 2.87 GHz, the NV ground state spin transition frequency, rather than the domain wall frequency. It is likely that our microwave stripline does not produce the same field strength at all frequencies. These results suggest that, similar to earlier studies of vortex gyrations \cite{Trimble2021}, DW driving results in off-resonant stray fields that will overlap with the NV transition frequency.

Additionally, extracting the Fourier component of the DW-generated microwave field in these simulations shows a strong enhancement; 0.1 mT driving fields result in a domain wall stray microwave field amplitude exceeding 3 mT, suggesting that bringing a DW and NV center into resonance can drastically reduce the $\pi$ pulse time. 

\section{\label{sec:level4}Conclusion}

In summary, we have shown that it is possible to detect localized GHz-scale DW dynamics in ferromagnetic systems using NV centers by deterministically placing them on top of a DW pinning site in Py nanowires. This detection, via enhanced spin-relaxation from the broadband stray fields generated, further expands the menagerie of magnetic dynamics that can be measured using NV centers. Additionally, the prospect of resonant NV-DW coupling opens up potential avenues for qubit control; in the microwave driven case, our simulations suggest that the presence of a proximal DW could increase the driving field at an NV center by over a factor of 30, drastically reducing the $\pi$ pulse time. Furthermore, this work opens up the possibility of driving NV centers using current driven DW oscillations, exploiting DWs as nanoscale microwave generators. 

\section{\label{sec:level5}Methods}

\subsection{\label{sec:level6}Domain Wall Oscillation Measurement}
We use standard electron beam lithography, electron beam thin film deposition, and a liftoff process to prepare 10 nm thick Py semicircular nanowires with a notch-shaped defect in the center. 100 nm diameter nanodiamonds (Adámas Nanotechnologies, 3 ppm NV) are placed using an AFM pick-and-place technique outlined elsewhere\cite{Rable2022}, yielding results like those shown in Fig.~\ref{fig:NDPlacement} (a), (c). MFM measurements, carried out in a Bruker ICON AFM with a low moment CoCr coated tip (Bruker, MESP-LM-V2), confirm nanodiamond proximity to a DW (Fig.~\ref{fig:NDPlacement} (b), (d)). DWs in the nanowires are created and destroyed using the shape anisotropy of the curved wires; they are nucleated by applying a magnetic field perpendicular to the wire at the defect site, and denucleated by applying a magnetic field tangential to the wire at the defect site. The in-plane nucleation field is applied using an N52 permanent magnet mounted on a highly repeatable stepper motor stage (Newport, ILS200PP Stepper Motor), and the denucleation field is applied using an N52 magnet positioned near the sample by hand. The nucleation magnetic field is calibrated in the plane of the sample using the NV centers in a single crystal CVD diamond film with known orientation; due to stage limits, the smallest field we can apply is 0.7 mT.

For optical polarization and readout of the NV center fluorescence, we use excitation via a 532 nm wavelength CW laser (Oxxius, LCX-532L-200-CSB-PP) and detection with an avalanche photodiode (ID Quantique, ID100). The laser is pulsed using an acousto-optical modulator (AOM) (Gooch and Housego, 15210) set up in a double pass configuration, scanned across the sample via a fast scanning mirror (Optics in Motion, FSM-101), and focused on the sample using a $100 \times 0.9$ numerical aperture (NA) objective lens (Nikon, $100 \times$ Plan Fluor $0.9$NA objective). During the measurements, we apply a microwave magnetic field , which drives both NV spin transitions and magnetization dynamics, via a microwave signal generator (Stanford Research, SG396), +43 dBm amplifier (Mini-Circuits, ZHL-16W-43-S+), and a 25 \textmu m diameter gold wire placed across the sample. 


\subsection{\label{sec:level7}Micromagnetic Simulations}

We perform micromagnetic simulations to design our nanowires and to eventually compare to our experimental results. These simulations are performed with the Mumax3 software package, which uses the Landau-Lifshitz-Gilbert equation:
	
\begin{equation*}
    \frac{\partial \vec{M}}{\partial t} = \gamma_{\text{LL}} \frac{1}{1+\alpha^2} (\vec{m} \times \vec{B_{\text{eff}}} + \alpha(\vec{m} \times (\vec{m} \times \vec{B_{\text{eff}}}))
    \tag{1}
    \label{eqn:LLG}
\end{equation*}

 \noindent to calculate the evolution of the magnetization $\vec{M}$ of finite ferromagnetic cells. In Eq. \ref{eqn:LLG}, $\alpha$ is the Gilbert damping of the material, $\gamma_{\text{LL}}$ is the gyromagnetic ratio of the material, and $B_{eff}$ is the effective magnetic field at that cell, which includes contributions from external, demagnetization, exchange, and anisotropy fields\cite{Vansteenkiste2014}.

The simulations use 5 nm x 5 nm x 10 nm cells with a saturation magnetization $M_{\text{s}}$ of 8  x $10^{5}$ A/m, an exchange constant $A_{\text{ex}}$ of 1.3 x $10^{-11}$ J/m, and a Gilbert damping parameter $\alpha$ of 0.0063, which are standard for Py.  

After defining the wire geometry, initial simulations of domain wall nucleation are performed by giving the system a constant (1,0,0) magnetization (pointing tangential to the wire direction at the defect site) with an applied magnetic field along the (0,1,0) direction. The system is then allowed to relax into a low energy state. At low fields, the magnetization continues to lie along the wire direction, but past a certain field, a transverse domain wall nucleates in the wire due to shape anisotropy. The magnetization of the nucleated state is used as the initial state for the domain wall dynamics simulations. In these simulations, the nucleated magnetization from the previous simulations is loaded, and then allowed to relax in an applied field, like before. 

To excite the system, we then apply a Gaussian pulse with a 20 ps full width half maximum (FWHM) and a 0.5 mT amplitude. The system is allowed to freely evolve in time for 20 ns and the average magnetization is sampled every 5 ps. Finally, we use a discrete fast Fourier transform (FFT) to analyze the data in the frequency domain, providing a resolution of 50 MHz and window from 0 to 100 GHz, well beyond the range of interest in this work.

For spatially resolved simulations, a shorter 1 ps FWHM pulse is used, and the magnetization is sampled in discrete 50 x 50 nm blocks composed of 100 cells. The average magnetization of each block is analyzed using an FFT as before and the image is composed of each block's FFT amplitude at the desired frequency. 

To determine the stray field produced by a domain wall, we drive the system at a constant microwave frequency for 20 ns and sample the magnetization of every cell within 500 nm of the pinning site every 5 ps. Then, treating every cell as an individual dipole, we solve for the magnetic field at the desired location by summing the individual dipole fields at each time point. An FFT can then be used to determine the microwave field's frequency distribution. 

\bmhead{Acknowledgments}

We acknowledge support from the University of Chicago, the U.S. Department of Energy Office of Science National Quantum Information Science Research Centers (Q-NEXT), and a seed grant from the Penn State Materials Research Institute. The authors would like to thank Eric Kamp for technical discussions and Michael Labella for sample fabrication advice. 

\section*{Declarations}

\begin{itemize}

\item Competing interests: The authors declare no competing interests.
\item Availability of data and materials: all raw data are avaiable upon request.
\item Authors' contributions: J.R. and N.S. conceived the idea for this project. J.R. and J. D. carried out the experiments and analyzed the data with input from N.S. who also supervised the project. All authors discussed the results. J.R. and N.S. wrote the manuscript with contributions from all authors.
\end{itemize}







\end{document}